# Topological dynamics and current-induced motion a skyrmion lattice

J C Martinez AND M B A Jalil

*Computational Nanoelectronics and Nano-device Laboratory*
*National University of Singapore, 4 Engineering Drive 3, Singapore 117576*

We study the Thiele equation for current-induced motion in a skyrmion lattice through two soluble models of the pinning potential. Comprised by a Magnus term, a dissipative term and a pinning force, Thiele's equation resembles Newton's law but in virtue of the topological character to the first two, it differs significantly from Newtonian mechanics and because the Magnus force is dominant, unlike its mechanical counterpart –the Coriolis force– skyrmion trajectories do not necessarily have mechanical counterparts. This is important if we are to understand skyrmion dynamics and tap into its potential for data-storage technology. We identify a pinning threshold velocity for the one-dimensional pinning potential and for a two-dimensional potential we find a pinning point and the skyrmion trajectories toward that point are spirals whose frequency (compare Kepler's second law) and amplitude-decay depend only on the Gilbert constant and potential at the pinning point.

The experimental discovery in 2009 of a hexagonal skyrmion lattice in MnSi under an external vertical magnetic field generated a convergence of efforts to understand better the interplay between the ferromagnetic exchange and Dzyaloshinskii-Moriya couplings in conjunction with crystalline field interactions in B20 compounds, i.e. magnetic materials lacking inversion symmetry (or chiral magnets)[1, 2]. A skyrmion is a planar topological spin texture whose spins are distributed in a circularly symmetric and continuous manner with the spin at the center pointing downward while all spins at the edge are pointing upward. Of particular interest for its application to information-storage technology, specifically the racetrack memory [3], is the effect of current on the magnetic texture since a relatively small current density is able to induce skyrmion motion, thus fueling hopes that ultra-low current densities might be feasible in the manipulation of magnetic structures [4, 5]. But this hope is not without fears since the mechanisms responsible for pinning and current-induced skyrmion motion are presently not well understood [6]. Moreover, experimentally, only very slow translation motion of skyrmions has been observed.

The standard tool of magnetization dynamics is the Landau-Lifshitz-Gilbert (LLG) equation

$$\left(\frac{\partial}{\partial t} + (\mathbf{v}_s \cdot \boldsymbol{\nabla})\right)\mathbf{M} = -\gamma \mathbf{M} \times \mathbf{B}^{\text{eff}} + \frac{\alpha}{M}\mathbf{M} \times \left(\frac{\partial}{\partial t} + \frac{\beta}{\alpha}(\mathbf{v}_s \cdot \boldsymbol{\nabla})\right)\mathbf{M} \qquad (1)$$

$\mathbf{B}^{\text{eff}} = -\frac{1}{\hbar\gamma}\frac{\delta \mathcal{H}}{\delta \mathbf{M}}$ is the effective field; $\gamma = \frac{g\mu_B}{\hbar} > 0$ the gyromagnetic ratio, $\mathbf{v}_s$ the spin velocity parallel to the spin current, $\mathcal{H}$ the free energy density, α the Gilbert damping constant, and β the coupling between the spin-polarized current and local magnetization due to nonadiabatic effects [7, 8]. An immediate consequence of Eq. (1) is that the magnitude of the magnetization $\mathbf{M}^2$ is conserved in time. The LLG equation has found application in a variety of systems such as ferromagnets, vortex filaments, and moving space curves and structures such as spin waves and solitons, to name a few [9]. As a time-dependent nonlinear equation, the LLG equation is arguable the most difficult equation to solve in theoretical physics. It is natural then to seek simplified versions of it. Since our interest here is on current-induced forces in a skyrmion lattice, we consider Thiele's simplification of it and treat the pinning mechanism phenomenologically [10]. Thiele projected the LLG equation onto the relevant translational modes [11] and in this way obtained an equation that could be regarded as a dynamical force equation but derived from a torque equation. To obtain it, we first assume a steady-state rigid texture $\mathbf{M} = \mathbf{M}(r - \mathbf{v}_d t)$, $\mathbf{v}_d$ standing for the skyrmion drift velocity. Then we cross multiply Eq. (1) by $\mathbf{M}$, followed by a scalar multiplication of the result by $\frac{1}{M}\partial_i \mathbf{M}$ and finally we integrate over the skyrmion area:

$$\mathbf{G} \times (\mathbf{v}_s - \mathbf{v}_d) + \overleftrightarrow{\mathcal{D}}(\beta \mathbf{v}_s - \alpha \mathbf{v}_d) = \nabla V, \qquad (2)$$

where $G_k = \frac{1}{4\pi}\varepsilon_{ijk}\int d^2r \frac{1}{M^3}\mathbf{M}\cdot\partial_i\mathbf{M}\times\partial_j\mathbf{M}$ is the dimensionless gryo-coupling vector and $\overleftrightarrow{\mathcal{D}}_{ij} = \frac{1}{2\pi}\int_{UC} d^2r \frac{1}{M^2}\partial_i\mathbf{M}\cdot\partial_j\mathbf{M}$ the dissipative dyadic. The gyro-term can be traced back to the Berry phase and pushes a moving a skyrmion perpendicularly to its direction of motion and is also referred to as the Magnus force [5, 11, 12]. The Magnus force is the counterpart of the Coriolis force in dynamics. The latter is a small correction to the dynamical equations [13], whereas the former, as we will see, dominates the dynamics of our skyrmion system. The Magnus force makes a spinning ball swerve one way as it passes the air; the Coriolis force is a fictitious force due to motion in moving noninertial frame. If we view Eq. (1) as an equation in the reference frame of the current, it seems more fitting to compare the first term with the Coriolis force. The dissipation term, which sums up the skyrmion's tendency toward a region of lower energy, originates from Gilbert damping. At the right-hand side we inserted a term due to a potential *V*, which models the pinning potential. Internal details of the skyrmion are ignored. Since the skyrmion is assumed to be perfectly rigid, it is not possible to deduce the pinning forces due to cancellation of forces for such a structure. Pinning is important not only in magnetics but also in superconductivity [14], soliton theory [15] and meteorology. For a skyrmion of winding number $Q$ = -1 [1], $\mathbf{G} \equiv g\hat{\mathbf{n}} = -4\pi\hat{\mathbf{n}}$, $\hat{\mathbf{n}}$ being the normal to the thin film and $\overleftrightarrow{\mathcal{D}}_{ij} = 5.577\pi\delta_{ij}$, $i = j = 1, 2$.

It is obvious that the Thiele equations are already a vast simplification over the original LLG equation. Nevertheless it still is nonlinear, albeit one involving only first-order derivatives. Unlike Newton's equations of motion, Thiele's equations are not time-reversal invariant. Moreover, the quantities $g$ and $\overleftrightarrow{\mathcal{D}}_{ij}$ are of topological origin in contrast with the dynamical parameters entering into Newton's equations. It is important then to gain familiarity with the Thiele equations if we are to understand current-induced motion in chiral magnets [16]. What is notable about this system is the topological character of the Magnus and dissipative parameters, a significant departure from mechanical systems. In the early 90s ideas about a topological quantum mechanics were in vogue [17]; we might now speak of a topological dynamics for the present system. In this paper, we present two models for which exact solutions of the Thiele equations can be derived. We find that these results are in excellent agreement with numerical results. Our findings allow us to identify key features of the dynamics. Insights from Newtonian mechanics do not necessarily translate into analogous situations for the Thiele case (for instance Kepler's second law does not hold in one model; in the other model Coriolis deflection occurs without forward motion).

We begin with a one-dimensional sinusoidal form for V: $V = -V_0 \cos(2\pi x/\lambda)$ and assume $\lambda$ much larger than skyrmion size [18]. Let us also assume constant spin current $\mathbf{v}_s$ in the x-direction so $v_{sy} = 0$. With $v_{sx}$ known, the drift velocities $v_x^d, v_y^d$ can be solved from

$$v_x^d = \frac{g^2 + \alpha\beta\mathcal{D}^2}{g^2 + \alpha^2\mathcal{D}^2} v_{sx} + V_0 \frac{\alpha\mathcal{D}}{g^2 + \alpha^2\mathcal{D}^2} \frac{2\pi}{\lambda} \sin\frac{2\pi x}{\lambda} \tag{3a}$$

$$v_y^d = \frac{g(\beta-\alpha)\mathcal{D}}{g^2 + \alpha^2\mathcal{D}^2} v_{sx} + V_0 \frac{g}{g^2 + \alpha^2\mathcal{D}^2} \frac{2\pi}{\lambda} \sin\frac{2\pi x}{\lambda} \tag{3b}$$

Since the solutions are translationally invariant, we take, for simplicity, the initial position to be the origin. Making use of the formula $\sqrt{A^2-1} \int \frac{dx}{A+\sin x} = 2\tan^{-1}\frac{1+A\tan\frac{x}{2}}{\sqrt{A^2-1}}$ we find

$$\tan\frac{\pi x}{\lambda} = r \frac{\sin\frac{\pi}{\lambda}V\sqrt{r^2-1}\,t}{\sqrt{r^2-1}\cos\frac{\pi}{\lambda}V\sqrt{r^2-1}\,t - \sin\frac{\pi}{\lambda}V\sqrt{r^2-1}\,t}, \quad y = \frac{g}{\alpha\mathcal{D}}(x - v_{sx}t) \tag{4}$$

in which $\mathcal{M} = \frac{g(\beta-\alpha)\mathcal{D}}{g^2+\alpha^2\mathcal{D}^2} v_{sx}$, $\mathcal{N} = \frac{g^2+\alpha\beta\mathcal{D}^2}{g^2+\alpha^2\mathcal{D}^2} v_{sx}$, $V = \frac{2\pi}{\lambda} V_0 \frac{\alpha\mathcal{D}}{g^2+\alpha^2\mathcal{D}^2}$, $U = \frac{2\pi}{\lambda} V_0 \frac{g}{g^2+\alpha^2\mathcal{D}^2}$ have dimensions of velocity whereas the ratio $r = \frac{\mathcal{N}}{V}$ is dimensionless. Equations (4) hold when $r > 1$. When $r < 1$, we must make the replacements $\sin \to i \sinh$, $\cos \to \cosh$ and $\tan \to i \tanh$. Since $-1 \leq \text{arctanh}\,\xi \leq +1$, x has a limit point when $r < 1$. This limiting point does not appear when $r > 1$.

There is another way to look at the case $r = 1$. The drift velocity is positive for all x provided $\frac{g^2+\alpha\beta\mathcal{D}^2}{g^2+\alpha^2\mathcal{D}^2} v_{sx} + \frac{2\pi}{\lambda} V_0 \frac{\alpha\mathcal{D}}{g^2+\alpha^2\mathcal{D}^2} \sin\frac{2\pi x}{\lambda} \geq 0$ so a threshold spin velocity $v_{\text{spin threshold}}$ is required: $v_{\text{spin threshold}} \geq \frac{2\pi}{\lambda} V_0 \frac{\alpha\mathcal{D}}{g^2+\alpha\beta\mathcal{D}^2}$. The case $r = 1$ corresponds to equality. For this case the second equation of Eq. (4) still holds but the first is replaced by $\frac{\pi x}{\lambda} = \cot^{-1}\left(\frac{\lambda}{\pi A t} - 1\right)$: one recognizes that pinning still occurs in this instance.

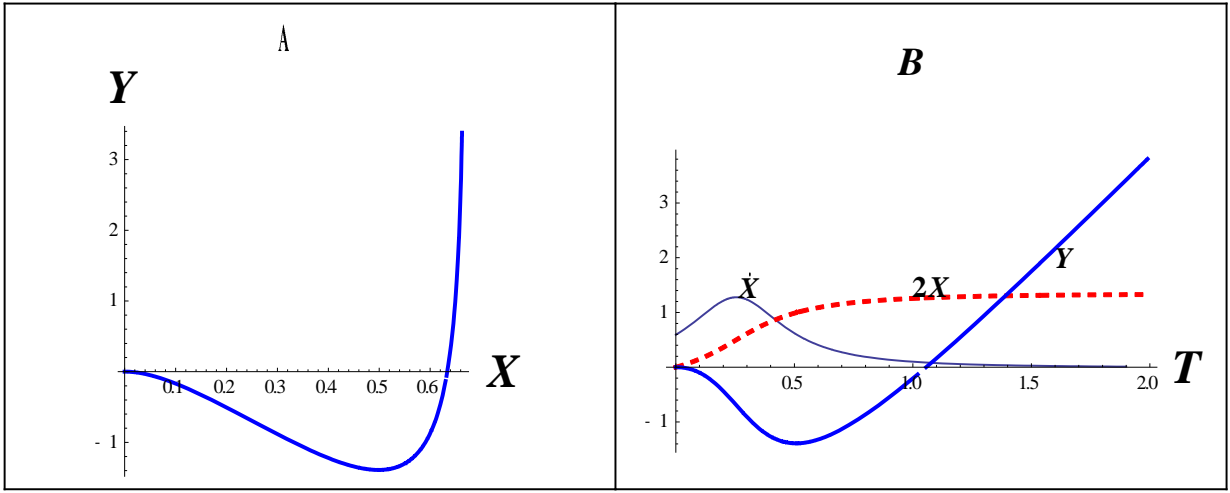

**Fig. 1** (a) Skyrmion trajectory for $v_s = 0.6$, $V_0 = 10$, $\alpha = 0.1$, $\beta = \alpha/2$ with the origin as starting point: $X = x/\lambda$, $Y = y/\lambda$. Numerical integration using *Mathematica* gives the same graph. (b) For the exact solution of Fig. 1(a) we display the position, $X(t)$, $Y(t)$, and velocity $\dot{X}$ as functions of time. For clarity we plot $2X(t)$ instead of $X$. See text for discussion.

Figure 1(a) displays the trajectory and Fig. 1(b) the position-time graphs for spin velocity $v_s = 0.6$ in the *x*-direction (this corresponds to $r < 1$) and $V_0 = 10$, $\alpha = 0.1$, $\beta = \alpha/2$. We use these latter parameters for Figs. 1 - 3. The $r = 1$ case for the parameters given corresponds to $v_s = 0.69037$. The starting point is always the origin. Equation (4) and numerical integration using *Mathematica* yield the same graphs. The first term of Eq. (2), which is the Magnus term, shows that the motion along the *y*-direction is due to the gyro-term and is large as comparison of the *X* and *Y* displacements on Fig. 1(b) indicates.

Figure 1(b) shows that the motion along the *x*-direction approaches a fixed or pinning point as the velocity $\dot{X}$ approaches zero asymptotically; whereas there continues to be a drift upward. We can think of the first term on the right-hand side of Eq. (3a) as the force component of the Magnus force opposed by the second term on the right-hand side, which represents a dissipative term, being proportional to $\alpha\mathcal{D}$. At the pinning point, these forces balance each other exactly. For Eq. (3b) the first term is the dissipative force, being proportional $g(\beta - \alpha)\mathcal{D}$, whereas the second term is the force component of the Magnus force in the y-direction. Close to the pinning point, this latter is much larger than the first.

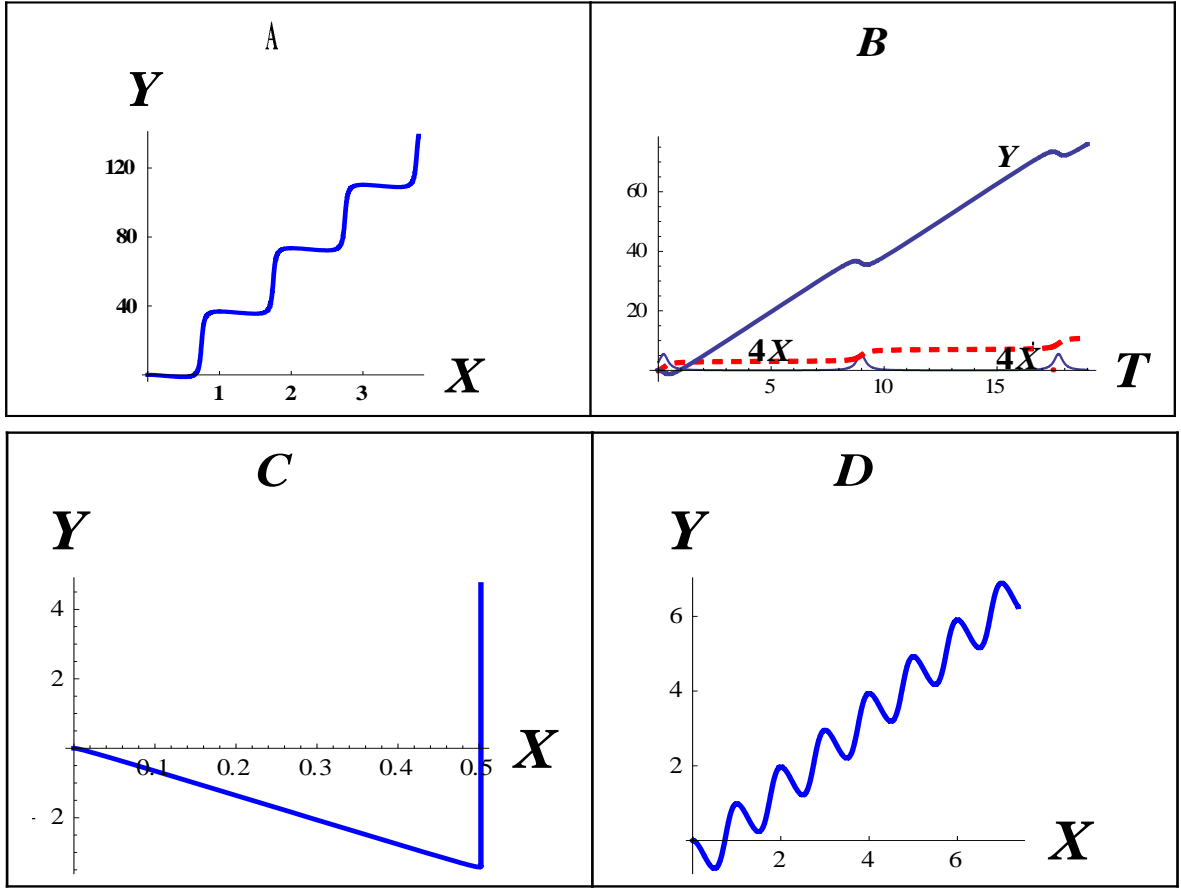

**Fig. 2** (a) and (b) The same as Fig. 1 but for $v_s$ = 0.7. In this case there is no pinning. Note the exaggerated scale for $X$ on Fig. 2(a). For greater clarity, we plot $2X$ and $2\dot{X}$. Only the first two 'steps' of Fig. 1(a) are noted in Fig. 1(b). (c). The same as Fig. 1(a) for $v_s$ = 0.01 and (d) for $v_s$ = 1.5

Figure 2(a), for spin velocity $v_s$ = 0.7 to the right, corresponds to $r > 1$. There is no pinning in this case. Both the exact and numerical solutions agree as before. The deflection upward by the Magnus force is large as Fig. 2(b) shows. It is due to the term linear in time in the equation for $Y(t)$ in Eq. (4); the first term, $UX/V$, is responsible for the small periodic

downward dips in the right-hand plot for $Y(t)$. Each of the almost horizontal steps in Fig. 2a corresponds to a crossing into the potential barrier (but there is no tunneling here).

Figures 2(c) and (d) show two instances quite removed from the $r = 1$ case. In the first where $v_s = 0.01$, the downward line is just the second half of Eq. (4), viz $y = \frac{g}{\alpha D}(x - v_{sx}t)$. For the upward drift the velocity is $-\frac{g}{\alpha D}v_{sx} > 0$, involves the topological quantities $g$ and $\mathcal{D}$. It is interesting to see upward motion *without* horizontal motion, in contrast with the Coriolis effect. Compared with Fig. 1a the trajectory is here shows sharp change of direction. In Fig. 2(d) with $v_s = 1.5$, the downward dips correspond to crossings into the pinning potential. The dips and upward movements are commensurate with each other unlike the case with Fig. 2(a). For large velocities this trend is maintained.

Consider next the attractive two-dimensional potential

$$V(x,y) = -U_0 \exp\left(-\frac{|x|}{\lambda} - \frac{|y|}{\lambda}\right). \tag{5}$$

This potential has cusps along the $x$ and $y$ axes. As above we assume a spin current only in the $x$-direction. The Thiele equations are

$$-gv_{dx} + \alpha \mathcal{D} v_{dy} = -gv_{sx} + \frac{U_0}{\lambda} e^{-\frac{|x|}{\lambda} - \frac{|y|}{\lambda}} \mathrm{sgn}(y) \tag{6a}$$

$$\alpha \mathcal{D} v_{dx} + gv_{dy} = \mathcal{D}\beta v_{sx} + \frac{U_0}{\lambda} e^{-\frac{|x|}{\lambda} - \frac{|y|}{\lambda}} \mathrm{sgn}(x) \tag{6b}$$

Let $\mathcal{N} = \frac{g^2 + \alpha \beta \mathcal{D}^2}{g^2 + \alpha^2 \mathcal{D}^2} v_{sx}$, $\mathcal{M} = \frac{g(\beta - \alpha)\mathcal{D}}{g^2 + \alpha^2 \mathcal{D}^2} v_{sx}$, and set $a = \frac{g}{g^2 + \alpha^2 \mathcal{D}^2}$, $\mathscr{b} = \frac{\alpha \mathcal{D}}{g^2 + \alpha^2 \mathcal{D}^2}$. Let $X = x/\lambda$, $Y = y/\lambda$, measure time with the same magnitude as $\lambda$, and define $\mathcal{U} = U_0/\lambda$. We find

$$\dot{X} = \mathcal{N} + (\mathscr{b}\,\mathrm{sgn}(X) - a\,\mathrm{sgn}(Y))\mathcal{U} e^{-|X|-|Y|}, \quad \dot{Y} = \mathcal{M} + (a\,\mathrm{sgn}(X) + \mathscr{b}\,\mathrm{sgn}(Y))\mathcal{U} e^{-|X|-|Y|} \tag{7}$$

Integrating we have

$$Y - Y_0 = \frac{a\,\text{sgn}(X) + \mathcal{b}\,\text{sgn}(Y)}{\mathcal{b}\,\text{sgn}(X) - a\,\text{sgn}(Y)}(X - X_0) + \left(\mathcal{M} - \frac{a\,\text{sgn}(X) + \mathcal{b}\,\text{sgn}(Y)}{\mathcal{b}\,\text{sgn}(X) - a\,\text{sgn}(Y)}\mathcal{N}\right)t. \qquad (8)$$

This holds for a given quadrant where signs of $X$ and $Y$ stay constant 'during' integration.

Again from the first of Eq. (7) we have $\dot{X}\text{sgn}(X) = \mathcal{N}\text{sgn}(X) + (\mathcal{b} - a\,\text{sgn}(X)\text{sgn}(Y))\mathcal{U}e^{-|X|-|Y|}$ and from the second, $\dot{Y}\text{sgn}(Y) = \mathcal{M}\text{sgn}(Y) + (a\,\text{sgn}(X)\text{sgn}(Y) + \mathcal{b})\mathcal{U}e^{-|X|-|Y|}$; adding and, staying in a fixed quadrant, we obtain $|\dot{X}| + |\dot{Y}| = \mathcal{N}\text{sgn}(X) + \mathcal{M}\text{sgn}(Y) + 2\mathcal{b}\mathcal{U}e^{-|X|-|Y|}$. In all cases below $\mathcal{N} > 0 > \mathcal{M}$. With $Z = |X| + |Y|$ we have finally

$$e^Z = e^{Z_0 + (\mathcal{N}\text{sgn}(X) + \mathcal{M}\text{sgn}(Y))t} + 2\frac{\mathcal{b}\mathcal{U}}{\mathcal{N}\text{sgn}(X) + \mathcal{M}\text{sgn}(Y)}\left(e^{(\mathcal{N}\text{sgn}(X) + \mathcal{M}\text{sgn}(Y))t} - 1\right) \qquad (9)$$

where $Z_0 = |X_0| + |Y_0|$. Equations (8) and (9) are the parametric equations of the trajectory for a *given* quadrant. For small times the trajectory is clearly straight. When $\frac{\mathcal{b}\mathcal{U}}{\mathcal{N}\text{sgn}(X) + \mathcal{M}\text{sgn}(Y)}$, which is proportional to α, is small we can also expect straight trajectories. In regions where this is large, we might expect curved trajectories.

Far from the origin the trajectory is straight. To see this, multiply the first equation of Eq. (7) by $Y$ and the second by $X$ and subtract. Far away from the origin we obtain $\frac{d}{dt}\tan^{-1}\frac{Y}{X} \to \frac{\mathcal{M}X - \mathcal{N}Y}{X^2 + Y^2}$, that is, if $\phi$ is the polar angle, then $\frac{d\phi}{dt} \to 0$ as $|X|, |Y| \to \infty$.

Figure 3(a) gives the trajectory for $\mathcal{U} = 40$ and $v_s = 0.0002$ (to the right) for the starting point S: (-6, 0). The plots of Eqs. (8) and (9) compares well with the *Mathematica* graph. Although the spin velocity $v_s$ is to the right, the direction of motion in the beginning (i.e., third) and following (second) quadrants are dictated by the gradient factor in Eq. (8),

$$\frac{a\,\text{sgn}(X)+b\,\text{sgn}(Y)}{b\,\text{sgn}(X)-a\,\text{sgn}(Y)} = \frac{g\,\text{sgn}(X)+\alpha\mathcal{D}\,\text{sgn}(Y)}{\alpha\mathcal{D}\,\text{sgn}(X)-g\,\text{sgn}(Y)}.$$ Note that this is a ratio of quantities of topological origin.

After the motion has entered into the first quadrant, time has now become large ($\sim 10^5$, see Fig. 3b) and the trajectory veers off only to assume a straight path outward to infinity. The effect of the cusps is evident and occurs only at the coordinate axes. These cusps might be suitable in modeling line defects.

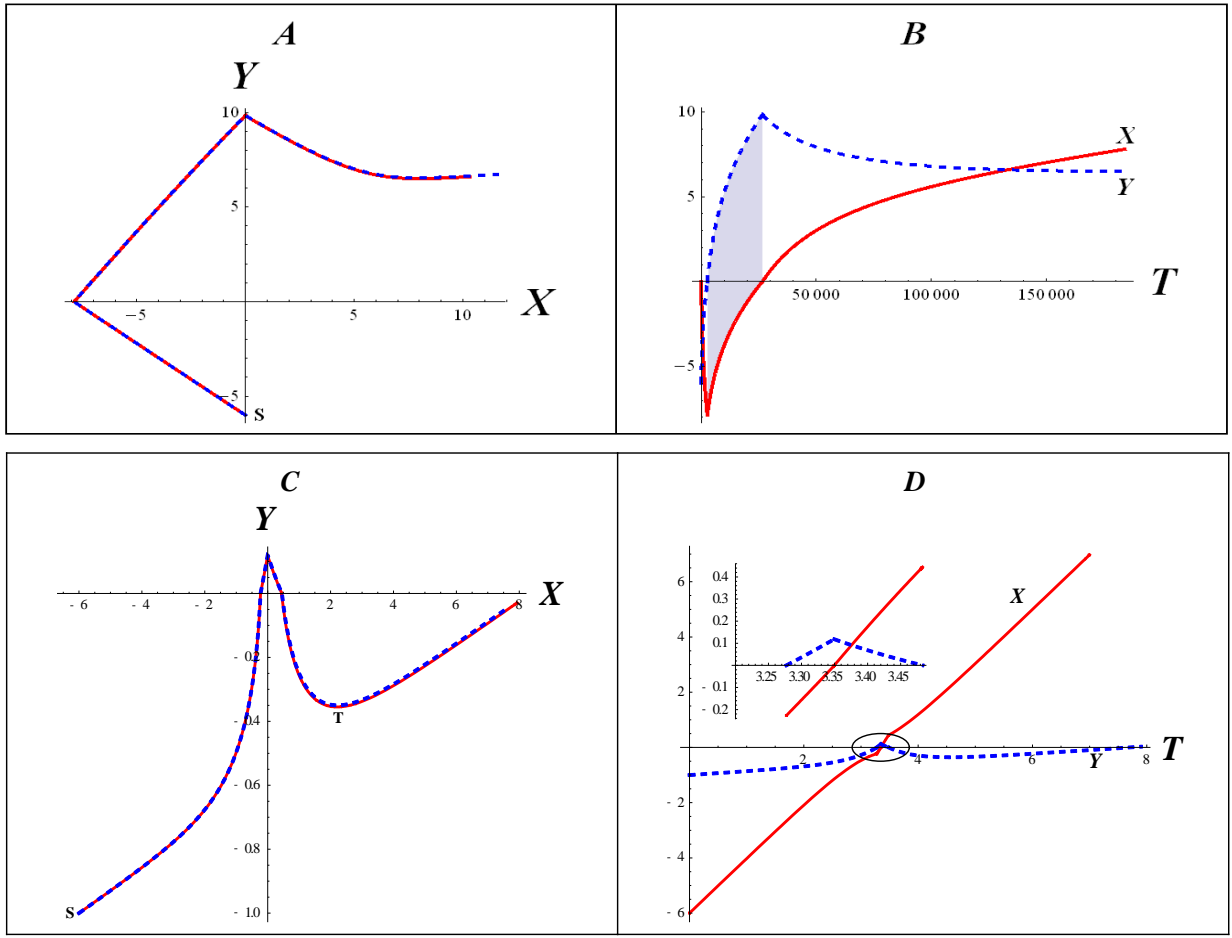

**Fig. 3** (a) Trajectory starting at S: (-6, 0) for $\mathcal{U} = 40$ and $v_s$ = 0.0002. The red curve is the exact result while the dashed is the one obtain via numerical integration with Mathematica. (b) Position-time graphs, $X(T)$, $Y(T)$. The discontinuities are due to the cusps. The shaded section corresponds to the second quadrant trajectory of Fig. 3(a) and is added for clarity. (c) The same as Fig. 3 (a) for S: (-6, -1) and $\mathcal{U} = 20$ and $v_s$ =2. (d) Position time graphs corresponding to Fig. 3(c). The inset is a zoom of the turn-around trajectory (circled) about the origin in Fig. 3(c).

Figure 3 (c) shows that trajectory from the starting point S: (-6, -1) for $\mathcal{U} = 20$ and $v_s$ = 2. As in Fig. 3(a), the exact and numerical results agree well with each other. The cusps are again evident. What appears striking here is the turn-around trajectory about the center of the potential at the origin. In fact a close-up of the trajectory around the origin, as shown in the inset of Fig. 3(d), indicates that the motion is very much like the first two parts of Fig. 3(a): they are straight-line segments whose gradients are given by the ratio $\frac{g\mathrm{sgn}(X)+\alpha\mathcal{D}\mathrm{sgn}(Y)}{\alpha\mathcal{D}\mathrm{sgn}(X)-g\mathrm{sgn}(Y)}$, which is of topological origin. Because the turn-around occurs much closer to the source of the potential than in Fig. 3(a) we see a faster reversal of the motion. At the point T in Fig. 3 (c), the drift is purely horizontal, i.e. the y-velocity vanishes. From the second equation of Eq. (7) we infer that this is where the dissipative (first term) force component is balanced by the Magnus (second) term.

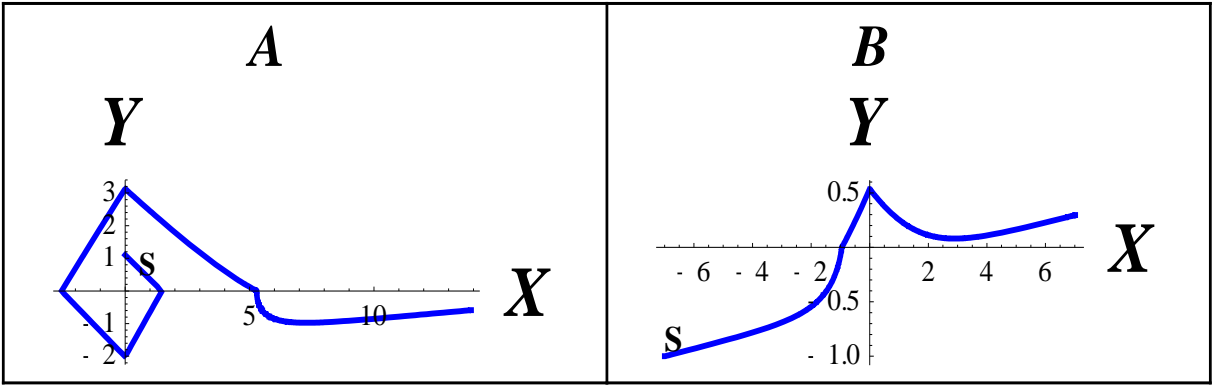

**Fig. 4** (a) Trajectories starting at S: (0, 1) for $\mathcal{U} = 30$ and $v_s$ = 0.01 and (b) S: (-7, -1) for $\mathcal{U} = 20$ and $v_s$ = 1.

Figure 4 shows other scenarios which indicate that whenever particles approach the origin they are bound to undergo the phenomenon already seen in Fig.3: straight-line trajectories whose gradients are given by the ratio of topological quantities $g$ and $\mathcal{D}$.

The LLG equation for the potential (5) does not have a pinning point even though the origin is clearly a local minimum. This is because of the cusps. To see pinning we take *two* identical attractive potentials $V(x, y)$, one centered at the origin as in Eq. (5), and another at (3.5, 0). We choose $\mathcal{U} = 2$, $v_{sx} = 0.03$ and consider the region $3.5 > x > 0, y < 0$, where the total potential is perfectly smooth. The Thiele equations take the form

$$X' = \mathcal{N} + (a + \ell\!\ell\,)\mathcal{U}e^{-X+Y} + (a - \ell\!\ell)\mathcal{U}e^{-3.5+X+Y} \tag{10a}$$

$$Y' = \mathcal{M} + (a - \ell\!\ell\,)\mathcal{U}e^{-X+Y} - (a + \ell\!\ell\,)\mathcal{U}e^{-3.5+X+Y} \tag{10b}$$

At a pinning point $(X_0, Y_0)$ both $X_0' = Y_0' = 0$. We can solve algebraically these equations with left-hand sides of Eq. (10) set to zero and obtain the pinning point: $X_0 = 1.825$ and $Y_0 = -0.6133$. (On the upper half, $Y > 0$, a similar calculation shows that there is no solution so we do *not* have pinning in the upper half plane.) Setting $X = 1.825 + x$, $Y = -0.6133 + y$ and introducing, $V = \mathcal{U}e^{-X_0+Y_0} = \mathcal{U}e^{-2.4383}$, $W = \mathcal{U}e^{-3.5+X_0+Y_0} = \mathcal{U}e^{-2.2883}$, Eqs. (10) can be expanded to first order:

$$x' = \mathcal{N} + (a + \ell\!\ell\,)V(1 - x + y) + (a - \ell\!\ell)W(1 + x + y)$$

$$y' = \mathcal{M} + (a - \ell\!\ell\,)V(1 - x + y) - (a + \ell\!\ell)W((1 + x + y)$$

These first-order equations can be easily solved exactly. The solutions have the time dependence factor $e^{\omega t}$, where $\omega \cong -\ell\!\ell(W + V) \pm 2i\sqrt{VW}$, since $\ell\!\ell \ll 1$. This describes a spiral with frequency $2\sqrt{VW}$ and amplitude decays in time through the factor $e^{-\ell\!\ell(W+V)t}$. The first result is a clear departure from Kepler's ($T^2 \propto r^3$) second law. From the definition of $\ell\!\ell$ we infer that the decay only depends on the Gilbert constant $\alpha$, but not on $\beta$. Moreover the frequency depends only on the strength of $V$ at the pinning point. The result is shown in Fig. 5 for $\mathcal{U} = 2$, $v_{sx} = 0.03$. The left-hand graph is obtained from Eqs. (10). The right-hand graph with starting point at (0.1, 0) is obtained by numerical integration via *Mathematica*. Equations (10) are applicable only in the smooth region $3.5 > x > 0, y < 0$.

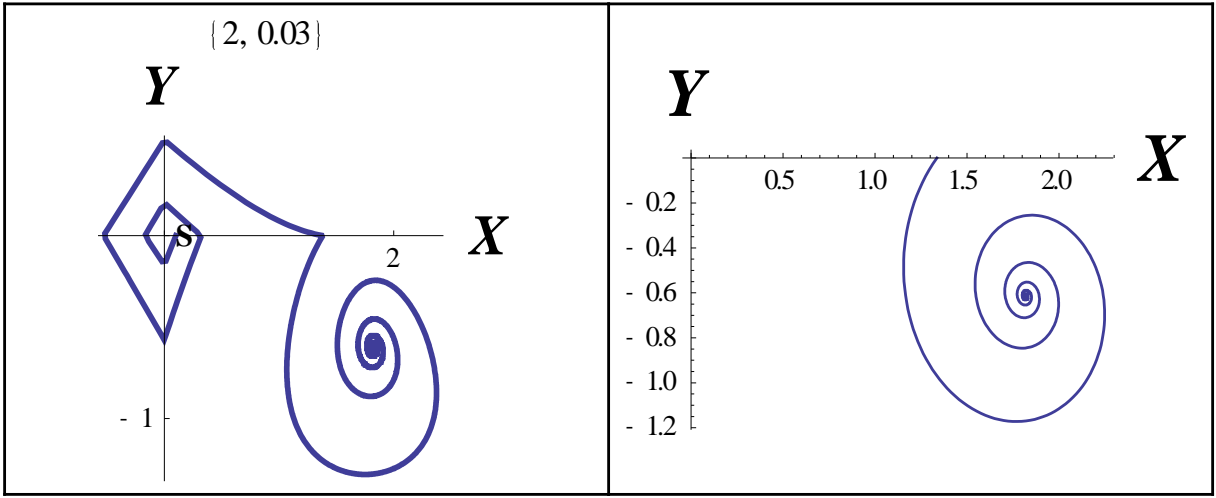

**Fig. 5** With two attractive potentials centered at the origin (0, 0) and at (3.5, 0), it is possible to pin a skyrmion. We choose $\mathcal{U} = 2$, $v_{sx} = 0.03$. On the right we have the result of Eq. (10). For comparison, on the left we use numerical integration with *Mathematica* from the starting point S = (0.1, 0). Note the effect of the cusps at the axes.

In summary we studied the Thiele equation for current-induced motion in a skyrmion lattice through two soluble models of the pinning potential. Thiele's equation is composed of a Magnus force, responsible for transverse motion relative to the current velocity, a dissipation force along the current velocity and the pinning force. The first two have topological origin whereas the third is imposed externally. In the first, one-dimensional model the Magnus force was found to dominate the dynamics and even transverse motion without corresponding skyrmion motion in the spin direction was possible. We saw a threshold velocity below which motion in the current direction is not allowed and which can be interpreted in terms of balance between the Magnus and the pinning forces. In the second two-dimensional case we saw the occurrence of straight trajectories in which the interplay of the Magnus and dissipative forces and, hence, their topological character are evident. Because of the peculiarities of the model, pinning onto a point was not possible; however with two potentials separated from each other, a pinning point could be found.  The trajectory close to the pinning point is a spiral whose frequency and amplitude decay depend only on the Gilbert constant and the strength of the potential at the pinning point. Kepler's second law did not hold in this system. We did not inquire into the effect of mass which can be a natural point for departure in the future [19].